\documentstyle[12pt,aasms4]{article}

\begin{document}

\title{Time-Delay Effect on the Cosmic Background Radiation by 
Static Gravitational Potential of Clusters}

\author{Da-Ming Chen\altaffilmark{1,2}, Xiang-Ping Wu\altaffilmark{2,3} 
        and Dong-Rong Jiang\altaffilmark{1,2}}

\altaffiltext{1}{Shanghai Astronomical Observatory, Chinese Academy
                 of Sciences, Shanghai 200030, China} 
\altaffiltext{2}{National Astronomical Observatories, Chinese Academy
                 of Sciences, Beijing 100012, China} 
\altaffiltext{3}{Beijing Astronomical Observatory, Chinese Academy
                 of Sciences, Beijing 100012, China} 

\begin{abstract}
We present a quantitative analysis of the time-delay effect 
on the cosmic background radiation (CBR) by
static gravitational potential of galaxy clusters. 
This is primarily motivated by growing observational evidence that
clusters have essentially experienced no-evolution
since redshift $z\approx1$, indicating that the contribution of 
a time-dependent potential to CBR anisotropy discussed in literature  
could be rather small for the dynamically-relaxed clusters.
Using the softened isothermal sphere model and 
the universal density profile for the mass distribution of rich clusters,
we calculate the CBR anisotropy by the time-delay effect and compare it 
with those generated by the thermal and kinematic S-Z effects 
as well as by the transverse motion of clusters.  
While it is unlikely that the time-delay effect is detectable in 
the current S-Z measurement because of its small amplitude of 
$10^{-6}$-$10^{-7}$ and its achromaticity, it nevertheless 
leads to an uncertainty of $\sim10\%$ in the measurement of 
the kinematic S-Z effect of clusters. Future cosmological application 
of the peculiar velocity of clusters to be measured through 
the S-Z effect should therefore take this uncertainty into account.
\end{abstract}


\section{Introduction}

The microwave sky behind a cluster of galaxies would be primarily distorted 
through the so-called Sunyaev-Zel'dovich (Z-S) effect -- 
the inverse Compton scattering
of the radiation by electrons in hot intracluster gas, which leads to
a temperature decrement of typically $\Delta T/T\sim10^{-4}$ in the 
Rayleigh-Jeans part of the spectrum for a cluster as rich as Coma.
The original goal of conducting the S-Z measurement is to estimate 
the Hubble constant $H_0$, while the recent attempt has also been made   
to determine the cluster baryonic (gas) mass  in combination with 
X-ray observation (Myers et al. 1997; Mason \& Myers 2000). Yet, 
the S-Z effect can only probe the distributions of intracluster gas, 
and provides no direct information about the underlying gravitational 
potential of the clusters, whereas the latter plays a much more important
role in cosmological study. This motivates us to address 
the following question: Can we simultaneously detect the gravitational 
imprint on the microwave sky by cluster potential during
the measurement of the S-Z effect ?

Essentially,  the gravitational potential of a cluster exercises an influence  
on the cosmic background radiation (CBR) through the gravitational
lensing effect and the Rees-Sciama effect (Rees \& Sciama 1968),
which consist primarily of three components: 
The first one arises from gravitational lensing, which  
alters the trajectories of the CBR photons, resulting
in a reduction of the small-scale intrinsic fluctuations in CBR 
(e.g. Kashlinsky 1988; Fukushige et al. 1996 and references therein). 
However, this component does not cause additional temperature variations 
in CBR because of the conservation of the CBR surface brightness; 
The second one comes from the Rees-Sciama effect, which accounts for 
the changing gravitational potential during the radiation
cross time, giving rise to a decrement of CBR temperature;
And the third component corresponds to the purely relativistic time-delay 
due to the deep gravitational potential, which enables us to receive 
the CBR photons emitted at an earlier time and thus makes a positive 
contribution to $\Delta T/T$. 
This last component is predicted by both gravitational
lensing and the Rees-Sciama effect.

In their pioneering work, 
Rees \& Sciama (1968) concluded that more significant contribution to
the CBR anisotropies could be due to the 
time-dependent potential other than the time-delay effect. 
This has led many authors to devote their interest to
the CBR anisotropies by the evolving gravitational 
potential of various nonlinear perturbations
(e.g. Seljak 1996 and reference therein).
Unfortunately, these results can hardly be applicable to galaxy clusters 
because of the unrealistic models for cluster matter distributions such
as the the Swiss cheese model (Dyer 1976) and 
the ``two step Vacuole" model (Rees \& Sciama 1968; Nottale 1984),
while numerical simulations are limited by the dynamical resolutions
on cluster scales especially inside core radii  
(Tuluie \& Laguna  1995; Seljak 1996). So far,
the only `plausible' constraint on the CBR temperature fluctuations
by the time-dependent potential of a rich cluster
may be the work by Chodorowski (1991). He studied the effect 
using a linear potential approximation and a pure spherical
infall model for cluster, and reached that 
$\Delta T/T\leq6.5\times10^{-7}$. In some sense,
such an estimate  can be regarded as an upper limit on the  CBR 
temperature fluctuations caused by the  time-dependent 
potential of a cluster.

On the other hand, numerous observations have claimed for 
a ``settled'' configuration of cluster matter evolution since $z\sim1$. 
It was shown more than a decade ago that optical counts of clusters 
are consistent with  no-evolution scenario for redshift out to at least 
$z\approx0.5$ (Gunn, Hoessel \& Oke 1986). The same conclusion
holds true for the X-ray selected clusters since $z\sim0.8$
(e.g. Fan, Bahcall \& Cen 1997; Rosati et al. 1998). 
Moreover, no significant differences in the
dynamical properties have been detected between high-redshift   
and low-redshift clusters, which include the X-ray luminosity,
the X-ray temperature, the velocity dispersion of cluster galaxies,
the mass-to-light ratio, the baryon fraction, etc. 
(Carlberg et al. 1996; Mushotzky \& Scharf 1997; 
Wu, Xue \& Fang 1999 and references therein). In particular,
the distribution of core radii of the intracluster gas in nearby
clusters is identical to that of distant ones ($z>0.4$) 
(Vikhlinin et al 1998).  Taking these observational facts as a whole, 
we feel that, in addition to the claim for a low-mass density
universe, the gravitational potential of clusters
is unlikely to have experienced a violent change since $z\approx1$.
Therefore, the CBR fluctuation due to the changing  
or time-dependent gravitational potential of  clusters at $z<1$ 
could be rather small. Therefore, we need probably concentrate on the 
`third component' and explore how large the time-delay effect on
the CBR fluctuations by the static gravitational well of a rich cluster 
would be.

The gravitational time-delay can be roughly estimated by 
$\Delta t_0=(2GM/c^3)\ln(4D_{cs}D_c/r_0^2)$ 
for a pointlike mass $M$, where $D_c$ is the distance to $M$, 
$D_{cs}$ is the separation between $M$ and the background source, 
and $r_0$ is the impact parameter.  
For a rich cluster with mass of $10^{15}M_{\odot}$ at cosmological distance
and CBR as the background source, $\Delta t_0\sim10^3$ - $10^{4}$ years. 
Recall that the time-delay between the images of the gravitationally
doubled quasar by a galactic lens of $10^{12}M\odot$
is $\sim1$ yr.  A time delay of $\Delta t_0=10^3$ - 10$^{4}$ yrs in the
observer's frame corresponds to a CBR temperature increment of 
$\Delta T_d/T\sim\Delta t_0/t_0\sim10^{-7}$ -- $10^{-6}$. 
Indeed, such an amount 
of temperature variation is about two orders of magnitude smaller than the
thermal S-Z effect. However, this value should be detectable with 
the future CBR detectors like MAP and Planck. 
Unlike the thermal S-Z effect and the kinematic S-Z effect due to 
the peculiar motion of galaxy clusters, 
$\Delta T_d/T$ arising from the time-delay effect is insensitive to 
the impact distance, and thereby may be the dominant contribution to 
the CBR temperature fluctuations at large radii from 
the cluster centers. Alternatively, the presence of the  time-delay 
component may complicate the measurement of the kinematic S-Z effect.
It has been suggested that the thermal and kinematic S-Z effects can be
separated using their different spectra especially at the frequency near
218 GHz where the thermal effect is zero (see Birkinshaw 1999).
Now, one may also need to subtract properly the contribution of the
time-delay effect in order to extract the kinematic effect although
the latter is still the dominant component.  
In particular, the time-delay effect by clusters may be comparable to the 
CBR perturbation caused by the transverse motion of the clusters as 
lenses (Birkinshaw \& Gull 1983; Gurvits \& Mitrofanov 1986). 
This will add further difficulty to the distinction between the two effects
even if the CBR measurements can reach a sensitivity of  
$\sim10^{-6}$--$10^{-7}$. On the other hand, a quantitative analysis of 
the time-delay effect by clusters will be helpful for our estimate 
of various uncertainties in the CBR measurements around clusters.

\section{Time-delay effect by clusters of galaxies}

CBR anisotropy by the Rees-Sciama effect is mainly associated with 
nonlinear and strongly evolving potentials, which can be well described 
by the following formalism (Martinez-Gonzalez, Sanz \& Silk 1990):
\begin{equation}
\frac{\Delta T}{T}=\frac{5}{3c^2}(\phi_o-\phi_e)-
    \frac{2}{c^2}\int_e^o d\vec{x}\cdot \nabla\phi+
    \vec{n} \cdot (\frac{\vec{v}_o}{c}-
                              \frac{\vec{v}_e}{c}),
\end{equation}
or equivalently,
\begin{equation}
\frac{\Delta T}{T}=\frac{1}{3c^2}(\phi_e-\phi_o)+
    \frac{2}{c^2}\int_e^o dt \frac{\partial \phi}{\partial t}+
    \vec{n} \cdot (\frac{\vec{v}_o}{c}-
                              \frac{\vec{v_e}}{c}),
\end{equation}
where the first term is  the Sachs-Wolfe effects, the second term
denotes the Rees-Sciama effect by a time-dependent potential well of 
nonstatic structure, and the third term simply represents the Doppler shift.
We are interested in the second term generated by an isolated 
structure like a galaxy cluster.   
For a static potential $\phi(\vec{r})$ embedded in the expanding universe,
we have
\begin{equation}
\phi(\vec{r})=-\int_V \frac{4\pi G \rho(\vec{r}^{\prime}) 
                 d^3\vec{r}^{\prime}}
          {\left|\vec{r}-\vec{r}^{\prime}\right|},
\end{equation}
in which $\vec{r}$ and $\vec{r}^{\prime}$ are connected to 
the comoving coordinates through $\vec{r}=a(t)\vec{x}$ and 
$\vec{r}^{\prime}=a(t)\vec{x}^{\prime}$, respectively. 
Because of the mass conservation of
$4\pi\rho(\vec{r}^{\prime}) d^3\vec{r}^{\prime}$,    
the partial derivative of $\phi$ with respective to $t$ reads
\begin{equation}
\frac{\partial \phi}{\partial t}=-\frac{\dot{a}}{a}\phi
\end{equation}
As a result, the CBR temperature anisotropy by the Rees-Sciama effect
of a static potential $\phi$ is 
\begin{equation}
\frac{\Delta T}{T}=-\frac{2}{c^3}\int\;\frac{\dot{a}}{a} \phi\; ds.
\end{equation}
where the integration is performed along the light  path $s$.
If we assume that the cosmological term $\dot{a}/a$ remains roughly
unchanged during the CBR photon cross time, i.e., 
the size of the nonlinear structure represented by
the static potential $\phi$ is relatively small, the above expression
can be written as  
\begin{eqnarray}
\frac{\Delta T}{T}=\frac{\dot{a}}{a}\Delta t;\\
\Delta t=-\frac{2}{c^3}\int\;\phi ds,
\end{eqnarray}
where $\Delta t$ is the relativistic time dilation due to the 
presence of $\phi$ in the framework of the linearized Einstein theory
(e.g. Cooke \& Kantowski 1975).
We can also view the problem from a different  
angle: The CBR photons can be trapped in the 
gravitational well $\phi$ and separated from the expansion of the 
Universe for a period of $\Delta t$. Namely, 
the CBR photons traveling through a  gravitational well
can conserve their energy with respect to the background photons.   
Defining the Hubble constant as $H(t)=\dot{a}/a$, we have from eq.(6) 
\begin{equation}
\frac{\Delta T_d}{T}=H(t)\Delta t=\sqrt{1+z_c}H_0\Delta t_0
\end{equation}
where the subscript $d$ denotes the time-delay component, and 
the time-delay in the observer's frame is
$\Delta t_0=(1+z_c)\Delta t$ with $z_c$ being
the redshift of the nonlinear structure,  Here and also hereafter
we assume a flat cosmological model with $\Omega_M=1$ and 
$H_0=50$ km s$^{-1}$ Mpc$^{-1}$.

We now focus on the numerical computation of the  
time-delay effect by clusters of galaxies. 
We approximate the mass distribution of clusters by two well-known models:  
the softened isothermal sphere model (hereafter SIS) 
and the cusped universal density profile  
(Navarro, Frenk \& White 1995; hereafter NFW):
\begin{eqnarray}
\rho_{SIS}=\frac{\sigma^2}{2\pi G}\frac{1}{r^2+r_c^2};\\
\rho_{NFW}=\frac{\delta_c \rho_{crit}}{(r/r_s)(1+r/r_s)^2}.
\end{eqnarray}
In SIS model,  $\sigma$ is the velocity dispersion of dark matter particles
and $r_c$ is the core radius, while in NFW profile, 
$\rho_{crit}\equiv 3H^2/8\pi G$ is the critical mass density for closure,
$\delta_c$ is the dimensionless characteristic density contrast, 
and $r_s$ is the scale length. 
To ensure the convergence of gravitational potential and the validity of
our assumption about the limited effective size of nonlinear structure,
we truncate the cluster at its virial radius defined by
\begin{equation}
M(r_{vir})=\frac{4\pi}{3}r_{vir}^3\Delta_c \rho_{crit},
\end{equation}
in which  $\Delta_c\approx200$ is the overdensity of dark matter 
halo with respect to $\rho_{crit}$. 
The gravitational potentials for these two models inside $r_{vir}$ 
are as follows: For SIS model
\begin{equation}
\phi_{SIS}=
       2\sigma^2\left[\frac{\arctan x}{x}+\frac{1}{2}\ln(1+x^2)-1-\frac{1}{2}
\ln(1+x_{vir}^2)\right], 
\end{equation}
where $x=r/r_c$ and $x_{vir}=r_{vir}/r_c$. For NFW
\begin{equation}
\phi_{NFW}=-4\pi Gr_s^2\rho_s\frac{\ln(1+x)}{x},
\end{equation}
where $\rho_s=\delta_c \rho_{crit}$ and $x=r/r_s$. Inserting these 
derived potentials into eq.(7) and performing the integration along
the light path across the cluster, we can obtain the CBR temperature 
fluctuations due to the static gravitational potential of clusters
approximated by SIS model and NFW profile, respectively.

\section{Comparison}

\subsection{S-Z effect}

For an isothermal $\beta$ model as the distribution of the hot plasma
inside a cluster, the thermal S-Z effect is (see Rephaeli 1995; 
Birkinshaw 1999)
\begin{eqnarray}
\frac{\Delta T_{{\rm TSZ}}(\theta)}{T_{\rm CBR}}=
          g(x_{\nu})y_0 \left[1+\left(\frac{\theta}{\theta_{x,c}}
          \right)^2\right]^{(1-3\beta)/2};\\
 g(x_{\nu})=\frac{x_{\nu}^2e^{x_{\nu}}}
          {(e^{x_{\nu}}-1)^2}\left(x_{\nu}\coth\frac{x_{\nu}}{2}-4\right),
\end{eqnarray}
where $x_{\nu}=h\nu/kT_{\rm CBR}$ is the dimensionless frequency, 
$T_{\rm CBR}=2.726$K is the present CBR temperature, and 
\begin{equation}
y_0=7.12\times 10^{-5}\frac{\Gamma(\frac{3\beta-1}{2})}
               {\Gamma(\frac{3\beta}{2})}
\left(\frac{n_{e0}}{10^{-3}{\rm cm}^{-3}}\right)
      \left(\frac{T_e}{10{\rm keV}}\right)
       \left(\frac{r_{x,c}}{\rm Mpc}\right),
\end{equation}
in which $n_{e0}$ and $T_e$ are the central electron number density
and temperature, respectively, and $r_{x,c}$ (or $\theta_{x,c}$) is 
the core radius for the $\beta$ model. The kinematic S-Z effect due to 
the peculiar motion $v$ of cluster along the line of sight is 
\begin{equation}
\frac{\Delta T_{\rm KSZ}(\theta)}{T_{\rm CBR}}=
          -\left(\frac{v}{c}\right)
          n_{e0}\sigma_{T}r_{x,c}\sqrt{\pi}
         \frac{\Gamma(\frac{3\beta-1}{2})}{\Gamma(\frac{3\beta}{2})}
         \left[1+\left(\frac{\theta}
         {\theta_{x,c}}\right)^2\right]^{(1-3\beta)/2}.
\end{equation}

There are two remarkable 
distinctions between the S-Z and time-delay effects: 
(1)Unlike the S-Z effect
which is independent of cluster redshift, the time-delay effect varies as  
$(1+z_c)^{3/2}$; 
(2)Both $\Delta T_{\rm TSZ}$ and  $\Delta T_{\rm KSZ}$ drop
sharply with the outward radius from cluster center, while 
$\Delta T_{d}$ is rather insensitive to cluster radius. 
In order to quantitatively compare the CBR temperature fluctuations 
arising from the thermal and kinematic S-Z effects and the time-delay effect,
we take a typical rich cluster at $z=0.1$, whose parameters are 
listed in Table 1, to proceed our numerical computations.  
For the NFW profile, we adopt the typical
values of $\alpha$($\equiv 4\pi G \mu m_p \rho_s r_s^2/kT_e$) and $r_s$
found from the fitting of the NFW expected X-ray surface brightness profiles
of clusters to the 
observed ones (e.g. Ettori \& Fabian 1999; Wu \& Xue 2000). 
The resultant CBR temperature variations $\Delta T(\theta)/T_{\rm CBR}$
are shown in Fig.1 for two different choices of the truncated cluster
radii: $r=r_{vir}$ and $r=10r_{vir}$.

\begin{deluxetable}{ccccccccc}
\tablecaption{Cluster parameters \label{table-1}}
\tablehead{
\colhead{ $\sigma$ (km/s)} & 
\colhead{$\beta$} & \colhead{$r_c$ (Mpc)} & 
\colhead{$n_{e0}$ (cm$^{-3}$)} & \colhead{$T_e$ (keV)} & 
\colhead{$v$ (km/s)} &\colhead{$\alpha$} & 
\colhead{$r_s$ (Mpc)}  
}
\startdata
  1100  & 2/3 & 0.25  & $3\times 10^{-3}$ & 7 & 500 & 
       10 & 0.8 \nl
\enddata
\end{deluxetable}

\placefigure{fig1}

It appears that 
for a typical cluster and an observing frequency $\nu=32$ GHz, 
the orders of magnitude of the maximum CBR temperature 
fluctuations by the thermal and kinematic S-Z effects and the time-delay 
effect are, respectively, $10^{-4}$, $10^{-5}$ and $10^{-6}$. Although
a temperature fluctuation of as low as $10^{-6}$ will be detectable
with the future space experiments like MAP and PLANCK, 
the fact that $\Delta T_{d}/T_{\rm CBR}$ is 
a slowly varying function of radius 
puts the actual measurement of the time-delay effect into a 
difficult position. One possible way is to measure the temperature
difference between two points separated by an angle $\Delta\theta$. 
The signature of 
$|\Delta T_d(\theta-\Delta\theta/2)-\Delta T_d(\theta+\Delta\theta/2)|$
could be identified if the S-Z effects can be removed.
Yet, the kinematic S-Z and time-delay  effects
may become to be indistinguishable as a result of  
the frequency-independent property, unless the CBR temperature profiles
with  a sensitivity of at least $10^{-7}$ can be obtained. 
In other words, the presence of the time-delay effect may yield 
an uncertainty of $\sim10\%$ in the measurement
of the central kinematic S-Z effect of clusters.

\subsection{Transverse motion of clusters}

In the Rayleigh-Jeans limit, the magnitude of the temperature fluctuation 
$\Delta T_{v}/T_{\rm CBR}$ due to the transverse motion of a cluster with
velocity $v$ can be estimated through (Birkinshaw \& Gull 1983;
Gurvits \& Mitrofanov 1986)
\begin{equation}
\frac{\Delta T_{v}(\theta)}{T_{\rm CBR}}\approx2
               \left(\frac{v}{c}\right)\delta(\theta),
\end{equation}
where $\delta(\theta)$ is the deflection angle produced by 
the projected cluster mass within $\theta$ along the line of sight.
For the SIS and NFW models, we have 
\begin{equation}
\delta_{SIS}(\theta)=4\pi\left(\frac{\sigma^2}{c^2}\right)
            \frac{\sqrt{\theta^2+\theta_c^2}-\theta_c}{\theta},
\end{equation}
and
\begin{equation}
\delta_{NFW}(\theta)=\frac{16\pi G\rho_s r_s^2}{c^2}
            \left(\frac{\theta_s}{\theta}\right) \times
\left\{
\begin{array}{ll}
\ln \frac{\theta}{2\theta_s}+\frac{\theta_s}{\sqrt{\theta_s^2-\theta^2}}
             \ln \frac{\theta_s+\sqrt{\theta_s^2-\theta^2}}{\theta}, & 
                  \theta < \theta_s;\\
\ln \frac{\theta}{2\theta_s}+\frac{\theta_s}{\sqrt{\theta^2-\theta_s^2}}
             \arctan \frac{\sqrt{\theta^2-\theta_s^2}}{\theta_s}, & 
                  \theta > \theta_s,
\end{array} \right.
\end{equation}
respectively, where $\theta_s$ is the angle distance of $r_s$. 
$\Delta T_{v}(\theta)/T_{\rm CBR}$ produces a two-sided pattern around
the moving cluster. Unlike the time-delay and S-Z effects, CBR will 
be unaffected by  the transverse motion of the cluster if we look through
its central region because of $\delta(0)=0$.
The maximum amplitude of $\Delta T_{v}/T_{\rm CBR}$ relative to 
$\Delta T_{v}(0)/T_{\rm CBR}$ for a typical rich cluster is
\begin{eqnarray}
\frac{\Delta T_{v}}{T_{\rm CBR}}\approx 0.9\times10^{-6} 
              \left(\frac{v}{10^3 {\rm km\;s^{-1}}}\right)
              \left(\frac{\sigma}{10^3 {\rm km\;s^{-1}}}\right)^2, & 
                        {\rm SIS};\\
\frac{\Delta T_{v}}{T_{\rm CBR}}\approx 1.1\times10^{-6} 
              \left(\frac{v}{10^3 {\rm km\;s^{-1}}}\right)
              \left(\frac{\alpha}{10}\right)
              \left(\frac{T}{7 {\rm keV}}\right), & 
                        {\rm NFW}.
\end{eqnarray}
Apart from their very different CBR patterns,  
$\Delta T_{\rm KSZ}/T_{\rm CBR}$,
$\Delta T_{v}/T_{\rm CBR}$ and $\Delta T_{d}/T_{\rm CBR}$ are all
achromatic, and the latter two effects also have the same order 
of magnitude. Consequently, the uncertainty in the measurement of
$\Delta T_{\rm KSZ}$ due to the combined effect of the time-delay 
and the transverse motion of clusters can become even larger than $\sim10\%$. 
It was suggested (Birkinshaw \& Gull 1983; Gurvits \& Mitrofanov 1986)
that the detection of $\Delta T_{v}/T_{\rm CBR}$ can be used as a method for
measuring the peculiar velocities of clusters. Such a motivation can now be 
complicated by the time-delay effect unless the detailed
patterns of the CBR anisotropies around clusters can be well mapped.

\section{Discussion and conclusions}

Indeed, the CBR temperature fluctuation caused by static gravitational 
potential of a rich cluster is very small,  
$\Delta T_{d}(\theta)/T_{\rm CBR}\sim10^{-6}$--$10^{-7}$, 
which is of 2--3 (1--2) 
orders of magnitude lower than the thermal (kinematic) S-Z effect,
but is nevertheless comparable to the effect produced by a transversely
moving cluster as the gravitational lens. The signals of the time-delay
effect and the transverse motion of clusters 
may remain to be indistinguishable from the kinematic S-Z effect
in current  S-Z measurement, unless one can acquire the detailed 
CBR temperature profile across clusters with 
a sufficiently high sensitivity of $\sim10^{-7}$. 
The presence of the time-delay effect and  
the transverse motion of clusters may lead to an uncertainty of $\sim10\%$ 
in the measurement of the kinematic S-Z effect 
due to the peculiar motion of clusters along the line of sight,
which gives a sense of how accurate and robust
one can use the kinematic S-Z measurement of clusters for
cosmological purpose. 

So far, 
we have not included the contribution from the time-dependent potential of
clusters. This is mainly based on the recent observations that 
the significant evolution of dynamical properties of clusters has not been
found since $z\approx1$. Therefore, our conclusion may not hold exactly true 
if clusters are still in the process of formation where the free infall  
plays a dominant role.  Our derived CBR temperature fluctuation due to
the time-delay effect can be comparable to that produced by
the time-dependent potential of clusters 
(e.g. Chodorowski 1991; Tuluie \& Laguna 1995; Tuluie, Laguna 
\& Anninos 1996). 
Nevertheless, in a similar way to the S-Z effect,
the $\Delta T/T_{\rm CBR}$ caused by the changing gravitational potential of 
a cluster shows a sharp drop along outward radius. So, it
would be possible to isolate the time-delay effect from the measured 
CBR fluctuations behind a cluster if one can have the high-sensitivity
CBR temperature profile.

Because of the unique property of the `long-distance' effect, i.e., 
$\Delta T_d/T_{\rm CBR}$ depends approximately on $\ln (1/\theta)$, 
one may worry about the issue of whether 
the time-delay effect by clusters can contaminate the global 
CBR power spectrum measured at small angles ($\sim10$--$100$ arcminutes) 
because clusters are a rare population in the Universe. Without a 
sophisticated statistical study of the effect on the CBR spectrum, 
our computation in the present paper suggests that 
the time-delay effect by clusters is unlikely to 
produce a noticeable contribution of
as high as few times $10^{-6}$ to the CBR anisotropies,  
which is compatible with the Rees-Sciama effect generated  
by large-scale matter inhomogeneities according to numerical simulations
(Tuluie \& Laguna 1995; Tuluie et al. 1996; Seljak 1996).  
Nevertheless, future precise CBR temperature measurements on smaller
angular scales should allow the time-delay effect of clusters 
to be included.

\acknowledgments
We gratefully acknowledge Tzihong Chiueh for useful discussion, and
the science editor of the journal, Prof. R. T. Vishniac, and 
an anonymous referee for  valuable
comments and suggestions. This work was supported by 
the National Science Foundation of China, under Grant No. 19725311.

\clearpage

\figcaption{Radial CBR temperature variations generated 
by the thermal (dotted lines) and kinematic (dashed lines) S-Z effects
and the time-delay (solid line) effect for a cluster at $z=0.1$.
For the thermal S-Z effect an observing frequency of $\nu=32$ GHz 
is assumed. Cluster properties are summarized in Table 1. 
Dependence of the effects on the truncated radii is shown for 
$r_{cut}=r_{vir}$  (upper panels) and  $r_{cut}=10r_{vir}$ (lower panels),
respectively.
\label{fig1}}


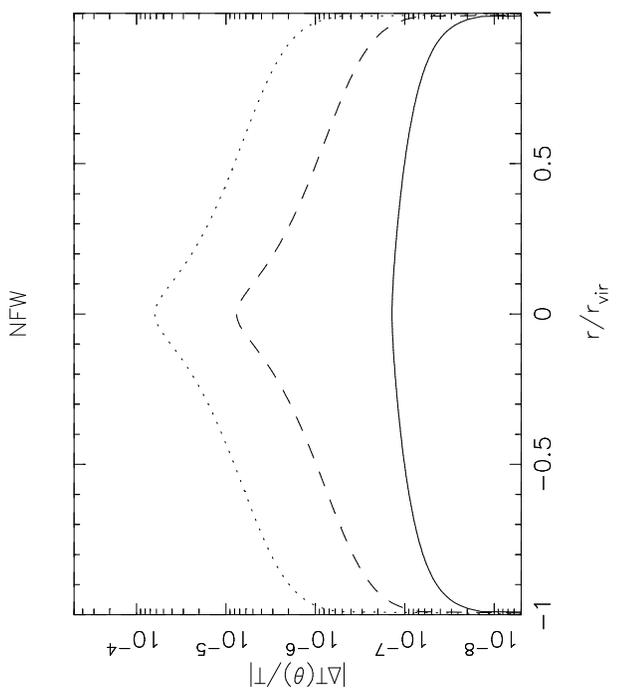
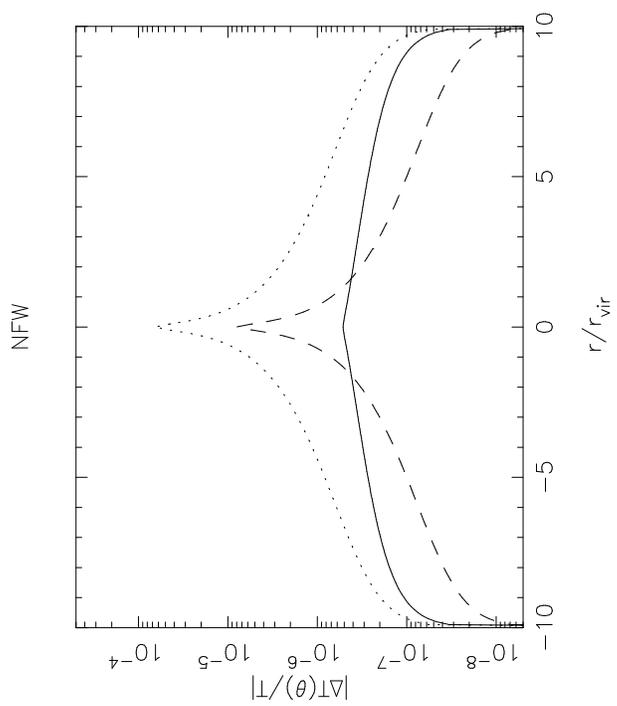
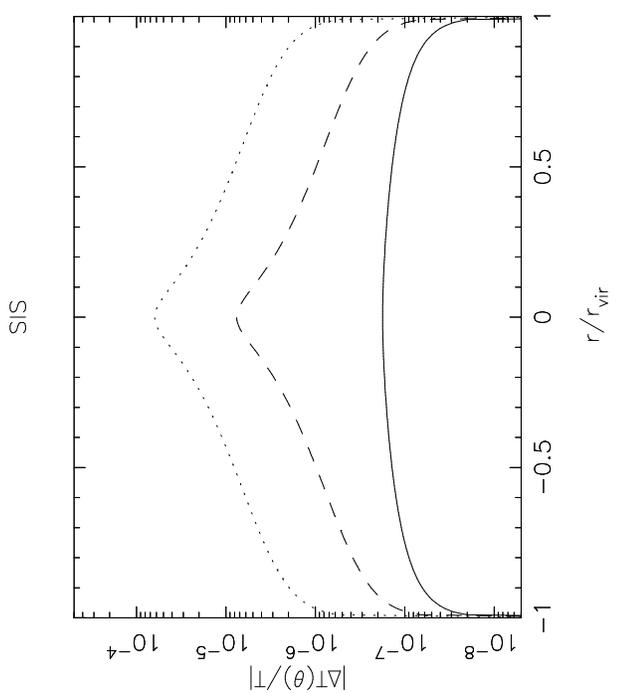
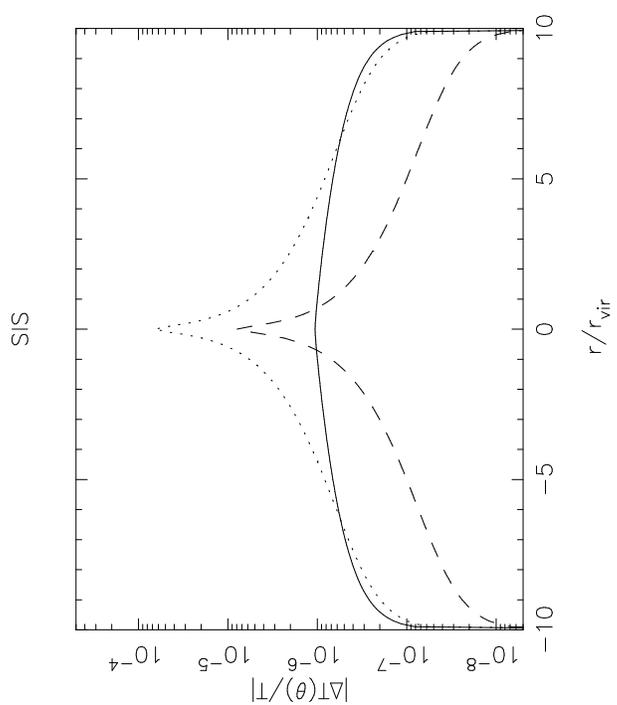
\end{document}